\documentclass[journal=langd5,manuscript=article,layout=twocolumn]{achemso}

\usepackage[version=3]{mhchem} 
\usepackage{setspace}

\author{Jonathan J. Harris}

\author{George A. Pantelopulos}

\author{John E. Straub} 
\email{jjh88@bu.edu; gpantel@bu.edu; straub@bu.edu}
\affiliation[Boston University]
{Department of Chemistry, Boston University, 590 Commonwealth Avenue, Boston, MA, 02215, USA}

\begin{onecolumn}

\title[Finite-size effects and optimal system sizes in simulations of surfactant micelle self-assembly]
  {Finite-size effects and optimal system sizes in simulations of surfactant micelle self-assembly}

\begin{document}
\begin{spacing}{1.2}
\begin{abstract}
The spontaneous formation of micelles in aqueous solutions is governed by the amphipathic nature of surfactants and is practically interesting due to the regular use of micelles as membrane mimics, for the characterization of protein structure, and for drug design and delivery. We performed a systematic characterization of the finite-size effect observed in single-component dodecylphosphocholine (DPC) micelles with the coarse-grained MARTINI model. Of multiple coarse-grained solvent models investigated using large system sizes, the non-polarizable solvent model was found to most-accurately reproduce SANS spectra of 100 mM DPC in aqueous solution. We systematically investigated the finite-size effect at constant 100 mM concentration in 23 systems of sizes 40 to 150 DPC, confirming the finite-size effect to manifest as an oscillation in the mean micelle aggregation number about the thermodynamic aggregation number as the system size increases, mostly diminishing once the system supports formation of three micelles. The accuracy of employing a multiscale simulation approach to avoid finite-size effects in the micelle size distribution and SANS spectra using MARTINI and CHARMM36 was explored using multiple long timescale 500-DPC coarse-grained simulations which were  back-mapped to CHARMM36 all-atom systems. It was found that the MARTINI model generally occupies more volume than the all-atom model, leading to the formation of micelles that are of a reasonable radius of gyration, but are smaller in aggregation number. The systematic characterization of the finite-size effect and exploration of multiscale modeling presented in this work provides guidance for the accurate modeling of micelles in simulations.

\end{abstract}

\section{Introduction} 
     The spontaneous self-assembly of surfactant molecules in aqueous solution is of great practical importance and fundamental interest due to the many applications of micelles and the forces which cause their formation. A significant challenge in the modeling of this process is capturing micelle size distributions, radii of gyration, and other characteristics that are in agreement with experiment, due in part to uncertainty in model predictions and variations in experimental observations. Dodecylphosphocholine (DPC), which has a dodecyl hydrocarbon tail and a zwitterionic phosphocholine head, is a commonly used surfactant. DPC micelles have been characterized in terms of ultracentrifugation,\cite{Lauterwein_Bosch_Brown_Wuthrich_1979}  dynamic light scattering\cite{Lauterwein_Bosch_Brown_Wuthrich_1979}, NMR\cite{Gao_Wong_1998,Gobl_Dulle_Hohlweg_Grossauer_Falsone_Glatter_Zangger_2010,Kallick_Deborah_A_Tessmer_Michael_R_Watts_Charles_R_Li_1995}, small-angle neutron scattering (SANS)\cite{Pambou_Crewe_Yaseen_Padia_Rogers_Wang_Xu_Lu_2015}, and small-angle X-ray scattering (SAXS)\cite{Lipfert_Columbus_Chu_Lesley_Doniach_2007,Oliver_Lipfert_Fox_Lo_Doniach_Columbus_2013}. Molecular dynamics (MD) simulations of DPC micelle self-assembly have been conducted with coarse-grained (CG)\cite{Marrink_Tieleman_Mark_2002,Marrink_de_Vries_Mark_2004,Sanders_Panagiotopoulos_2010} and all-atom (AA) modeling with explicit\cite{Urano_2018,Abel_Dupradeau_Marchi_2012,Urano_2019,Yoshii_2018,Takeda_2019,Zhang_2019} and  implicit\cite{Lazaridis_Mallik_Chen_2005,Mori2020,Chen2019} solvent. Fundamental aspects of the mechanism of self-assembly, the equilibrium state of micellar solutions, and the impact of micelle encapsulated impurities on micelle size, however, remain poorly understood.

     A special consideration in designing micelle simulations with periodic boundary conditions (PBC) is the size of the system, specifically the number of surfactants present in one unit cell. In complex lipid bilayer mixtures, it has been shown that the liquid-liquid phase separation transition critically depends on system size\cite{George_2017,Huang_1993}. In their self-assembly work with dissipative particle dynamics (DPD), Johnston \latin{et al.} consider an adequate system size to be one which contains a sufficient number of surfactants for the formation of five micelles\cite{Johnston_Swope_Jordan_Warren_Noro_Bray_Anderson_2016}. Kindt predicted the impact of finite-size on the aggregation number distribution when the total number of surfactants present is significantly smaller than the thermodynamic limit\cite{Kindt_2013,Zhang_Patel_Beckwith_Schneider_Weeden_Kindt_2017}. According to Kindt, even systems with as few as one or two micelles can potentially have size distributions that approximate what is expected in the thermodynamic limit\cite{Kindt_2013}. Because of the microsecond-order timescales associated with surfactant aggregation\cite{Marrink_de_Vries_Mark_2004}, it is of computational interest to determine the smallest possible system size that will produce a realistic micelle size distribution.

     In Tanford’s treatment of micelle self-assembly, the set of aggregation numbers of a micellar system is governed by forces that promote and limit aggregation\cite{Tanford_1973}. In particular, aqueous micelle formation is promoted by the tendency of the hydrocarbon tails to associate with each other, rather than with water. It is limited, in the case of zwitterionic DPC, by the comparatively stable solvation of the head groups as free monomers\cite{Tanford_1973}. The preferred aggregation number of the micellar system occurs at the size that optimally balances the promoting and limiting forces, resulting in a unimodal and approximately gaussian size distribution\cite{Tanford_1974}.
     
     Aggregation numbers for DPC micelles have been determined by various experimental methods. The particle weight obtained by ultracentrifugation of a 20 mM DPC solution by Lauterwein et al. was used to derive an aggregation number of 56 $\pm$ 5 surfactants per micelle\cite{Lauterwein_Bosch_Brown_Wuthrich_1979}. Kallick et al. used NMR experiments to obtain an aggregation number of 44 $\pm$ 5 surfactants for DPC at 228 mM\cite{Kallick_Deborah_A_Tessmer_Michael_R_Watts_Charles_R_Li_1995}. SANS experiments with 100 mM DPC conducted by Pambou \latin{et al.} resulted in an aggregation number of 70.6 $\pm$ 5 surfactants\cite{Pambou_Crewe_Yaseen_Padia_Rogers_Wang_Xu_Lu_2015}. SAXS methods have been used to obtain a range of aggregation numbers for DPC between 68 and 80 at 77 mM surfactant concentration\cite{Oliver_Lipfert_Fox_Lo_Doniach_Columbus_2013}. The large observed differences in aggregation number appear to result from the differences in surfactant concentration, as well as the indirect nature of observation, and the sensitivity of the models used in data analysis.

    Simulation studies of CG and AA DPC micelle formation have also been performed with not only varying surfactant concentrations, but also with different system sizes. In the AA single micelle self-assembly simulations conducted by Abel \latin{et al.}, 54 DPC surfactants were simulated from random initial configuration at a concentration of 200 mM\cite{Abel_Dupradeau_Marchi_2012}. 54 DPC were chosen based on the Lauterwein experiments\cite{Lauterwein_Bosch_Brown_Wuthrich_1979}, as well as proton NMR experiments\cite{Gao_Wong_1998,Gobl_Dulle_Hohlweg_Grossauer_Falsone_Glatter_Zangger_2010,Kallick_Deborah_A_Tessmer_Michael_R_Watts_Charles_R_Li_1995}, which reported similar aggregation numbers. Marrink et al. simulated 54 DPC at 460 mM and 120 mM, finding that a wormlike micelle formed at the higher concentration, and a spherical micelle formed at the lower concentration, both with an aggregation number of 54 surfactants\cite{Marrink_Tieleman_Mark_2002}. In MARTINI simulations of 400 DPC at 40 mM, Marrink \latin{et al.} reported a range of aggregation numbers from 40 to 70, but noted that the simulation had not fully converged after 1 $\mu$s\cite{Marrink_de_Vries_Mark_2004}. In a MARTINI self-assembly simulation with 175 DPC at 370 K and 126 mM, Sanders \latin{et al.} found an equilibrium size distribution that was unimodal and centered around 45 surfactants\cite{Sanders_Panagiotopoulos_2010}. 

     In order to efficiently simulate micelles, it is of interest to determine the lower limit of the number of surfactants necessary for a self-assembly simulation to produce a realistic size distribution. In 1996, Palmer \latin{et al.} simulated micelle self-assembly using a CG model with 100 surfactants. They reported multi-modal size distributions with sharp, separated peaks\cite{Palmer_Liu_1996}, suggesting the presence of a finite-size artefact. Kindt predicted that, in the region of small numbers of surfactants, the mean number of micelles and mean aggregation number oscillate around the values expected at the thermodynamic size limit, at which the number of micelles varies linearly with the total number of surfactants in the system\cite{Kindt_2013}. Although this finite-size effect has been predicted in other work\cite{Kindt_2013,Zhang_Patel_Beckwith_Schneider_Weeden_Kindt_2017}, no study has been conducted to fully characterize the effect on self-assembly simulations over an interval of system sizes approaching the thermodynamic limit.
 
    In this work, we extensively characterize the finite-size effect on the aggregation of DPC surfactants using the CG MARTINI 2 model. Using the SANS spectra of 100 mM DPC micelles\cite{Pambou_Crewe_Yaseen_Padia_Rogers_Wang_Xu_Lu_2015} of Pambou \latin{et al.} as a reference\cite{Pambou_Crewe_Yaseen_Padia_Rogers_Wang_Xu_Lu_2015}, we evaluated the ability of four MARTINI 2 solvent models to accurately simulate DPC micelle self-assembly. We determine the non-polarizable solvent model to produce the most accurate SANS spectrum. We then investigate the finite-size effect on micelle size distributions with this model at 100 mM surfactant concentration by simulating systems of 40 to 150 surfactants, at an interval of 5 surfactants, resulting in the formation of 1, 2 and 3 micelles throughout this range. We observe oscillations in the preferred aggregation numbers and number of micelles across the different system sizes, similar to the predictions of Kindt\cite{Kindt_2013}. We find that the finite-size effect is largely diminished once there are sufficient surfactants for three micelles to form at an aggregation number consistent with what is expected in the thermodynamic limit. The micelle size distributions observed for each increasing system size are analyzed by using the thermodynamic model of Tanford\cite{Tanford_1974}. We also test the efficacy of employing a multiscale approach by using large 500-DPC MARTINI simulations to simulate micelle self-assembly, and then back-mapping with the backward.py method\cite{Wassenaar_2014} to CHARMM36 AA representations. The CG and back-mapped AA simulations are compared directly to an experimental SANS spectrum\cite{Pambou_Crewe_Yaseen_Padia_Rogers_Wang_Xu_Lu_2015}. We find that the back-mapped AA micelles shrink to smaller than expected sizes due to the significantly larger volume of each MARTINI DPC molecule, suggesting that MARTINI 2 size distributions are significantly smaller than CHARMM36 at equilibrium, which indicates that the back-mapped systems are not in equilibrium. 
    
    Taken together, our results provide a consistent picture of DPC surfactant self-assembly at a range of system sizes, evaluate the accuracy of the MARTINI model for micelle simulation, and establish appropriate system sizes for the study of DPC micelles, which may be generalized as a method for determining the lower limit of the number of a variety of surfactants for self-assembly.
     
\section{Methods}

\subsection{A. Identifying the most accurate MARTINI model for micelle self-assembly}

     Four systems of the same composition were built with different versions of the MARTINI 2 force field for water and ions: nonpolarizable water (W)\cite{Marrink_Risselada_Yefimov_Tieleman_de_Vries_2007}, polarizable water (PW)\cite{Yesylevskyy_Schafer_Sengupta_Marrink_2010}, refined polarizable water (refPOL)\cite{Michalowsky_Schafer_Holm_Smiatek_2017}, and refPOL with polarizable ions (polIon)\cite{Michalowsky_Zeman_Holm_Smiatek_2018}. All of the systems used the v2.0 force field for DPC\cite{Marrink_Risselada_Yefimov_Tieleman_de_Vries_2007,Marrink_de_Vries_Mark_2004}, consisted of approximately 100 mM DPC with 150 mM NaCl, and contained 500 DPC molecules, 750 Na$^{+}$ ions, 750 Cl$^{–}$ ions and 68775 MARTINI waters. In the W system, 10\% of the water molecules were WF antifreeze. The choice of 500 DPC molecules was made in an effort to avoid finite-size effects. We expected that the systems would behave as expected in the thermodynamic limit. Six replicates of each system were simulated from random initial starting configurations. The W systems were simulated for 5.2 $\mu$s and PW, refPOL, and polIon were simulated for 4.0 $\mu$s. The MARTINI W water model was determined to be the most appropriate model for DPC micelle self-assembly as it led to the best agreement with the experimental SANS spectrum\cite{Pambou_Crewe_Yaseen_Padia_Rogers_Wang_Xu_Lu_2015}. In further simulations investigating finite-size effects on micelle formation, the W model was employed. 

\subsection{B. CG systems for probing finite-size effects}
     
     To characterize this finite-size effect, systems with $N$ = 40 to 150 DPC were constructed, at an interval of 5 surfactants. In order to obtain reliably averaged data, 20 replicates of each system were simulated from random initial starting configurations. The length of each simulation varied depending on the system size, and was adjusted to achieve an equilibrium micelle size distribution. System sizes $N$ = 40 to 100 were simulated for 2 $\mu$s, while system sizes N = 105 to 150 were simulated for 5 $\mu$s.  In each system, the numbers of water beads and NaCl ions were adjusted so that the concentration of DPC for each system was held to approximately 100 mM and the concentration of NaCl was held to approximately 150 mM. 
     
\subsection{C. CG to AA back-mapping for micelle characterization}

    To test the efficacy of a multiscale approach to micelle simulation, DPC micelles were self-assembled using the MARTINI model with non-polarizable solvent (W) and 500 surfactants. The resulting equilibrium configuration was back-mapped to the AA CHARMM36 force field. The differences between the micelles in the two representations were evaluated in terms of radius of gyration, size distributions, preferred aggregation numbers, and SANS spectra, along with the experimental SANS spectrum\cite{Pambou_Crewe_Yaseen_Padia_Rogers_Wang_Xu_Lu_2015}. The AA systems were constructed by back-mapping the W systems with $N$ = 500 DPC using backward.py\cite{Wassenaar_2014} and the DPPC.map file which was modified to suit DPC. The resulting AA systems contained 500 DPC, 275100 waters, and 750 each sodium and chloride ions.  This proportion of species corresponds to a DPC concentration of approximately 100 mM, in concurrence with the experimental conditions\cite{Pambou_Crewe_Yaseen_Padia_Rogers_Wang_Xu_Lu_2015}. All 6 of the MARTINI W systems were back-mapped for DPC atom positions only from their 5200 ns configurations and simulated for an additional 20 ns with the AA representation.
    
\subsection{D. CG model simulation details} 

    The MARTINI model test systems were composed using the GROMACS 2018.3 insert-molecules function, which randomly places molecules in the simulation box. Dodecahedral periodic boundary conditions were used. All of the MARTINI systems were energy minimized, equilibrated, and performed at constant NPT using GROMACS 2018.3 on GPUs\cite{Abraham_Murtola_Schulz_Pall_Smith_Hess_Lindah_2015,Pronk_Pall_Schulz_Larsson_Bjelkmar_Apostolov_Shirts_Smith_Kasson_van_der_Spoel_et_al_2013}. The systems were energy minimized using steepest descent. For MD, the leap-frog integrator was employed using a 20 fs time step. The Verlet cutoff-scheme for neighbor searching was applied. For non-bonded interactions, Particle-Mesh Ewald (PME) electrostatics was employed over a range of 0 to 1.1 nm with an ‘epsilon-r’ relative dielectric constant of 2.5 for all systems except for W, which had a dielectric constant of 15. Lennard-Jones was applied with a shifting function from 0.9 nm to the cutoff at 1.1 nm. For the thermostat, velocity-rescaling was employed with a coupling time of 1 ps at 295 K. Isotropic Parrinello-Rahman pressure coupling was used at 1 bar with a coupling time of 12 ps and 4.5 $\times$ 10$^{–5}$ bar$^{–1}$ compressibility. Bead coordinates and system energies were written every 1 ns. 

  The systems designed to extensively investigate finite-size effects used the same parameters described above for the MARTINI W test system. The particle definitions for water, ions, and DPC correspond to the MARTINI v2.0 force field\cite{Marrink_Risselada_Yefimov_Tieleman_de_Vries_2007,Marrink_de_Vries_Mark_2004}.

\subsection{E. AA model simulation details}
  
     For the AA system, the CHARMM36 force field\cite{Allouche_2012a,Soteras_Gutierrez_Lin_Vanommeslaeghe_Lemkul_Armacost_Brooks_MacKerell_2016,Yu_He_Vanommeslaeghe_MacKerell_2012a,Yu_He_Vanommeslaeghe_MacKerell_2012b} in was used for all particle definitions with CHARMM36 TIP3P water. After back-mapping the DPC positions, the systems were solvated as described above using insert-molecules and simulated using GROMACS 2020 with GPUs\cite{Abraham_Murtola_Schulz_Pall_Smith_Hess_Lindah_2015,Pronk_Pall_Schulz_Larsson_Bjelkmar_Apostolov_Shirts_Smith_Kasson_van_der_Spoel_et_al_2013}. The systems were minimized using the steepest descent minimization algorithm.  For MD, the leap-frog integrator with a 2 fs time step was used. The ‘Verlet’ cutoff-scheme for neighbor searching was applied with updates every 20 steps. For non-bonded interactions, Particle-Mesh Ewald (PME) electrostatics from 0 to 1.2 nm and ‘Shift’ Lennard-Jones Van der Waals from 1.0 to 1.2 nm were used. For the thermostat, velocity-rescaling was used with a coupling time of 1 ps at 295 K. Isotropic Parrinello-Rahman pressure coupling was used at 1 bar with a coupling time of 2 ps and 4.5 x 10-5 bar-1 compressibility. Atom coordinates and system energies were written every 100 ps.
 
\subsection{F. Thermodynamic analysis using Tanford model}
     The DPC molecules were clustered into micelles so that the equilibrium size distribution could be constructed and characterized. For all simulations in this study, single-link hierarchical clustering was used with a cutoff of 9 Å for the MARTINI systems and 12 Å for the all atom systems. A micelle was defined as a cluster with more than 10 surfactants. The cutoffs were determined based on visual inspection, radial distribution functions (for CG), radius of gyration (for AA), and the number of clusters present over time (see SI). The micelle definition was determined based on the populations of the size distributions, which generally do not contain clusters between 3 and 10 surfactants at equilibrium, no matter the system size. The clustering and micelle characterization was conducted using python with Scipy and MDAnalysis libraries\cite{Gowers_Linke_Barnoud_Reddy_Melo_Seyler_Domanski_Dotson_Buchoux_Kenney_et_al_2016,Michaud_Agrawal_Denning_Woolf_Beckstein_2011}. All MD visualizations and snapshots were made using Visual Molecular Dynamics (VMD)\cite{William_Humphrey_Andrew_Dalke_1996}.

        The time to reach equilibrium for each CG self-assembly trajectory was determined based on the time series of the number of clusters present in each system over time. In order to quantify the point at which each trajectory had converged into a stable micelle size distribution, the ergodic measure\cite{Kirkpatrick1989,Mountain1990,Straub809},
\begin{equation}
\Omega(t) = \left[ \left( \frac{1}{t} \sum_{i=0}^t n(i) \right) - \left( \frac{1}{T} \sum_{i=0}^T n(i) \right) \right]^2
\label{equation1}
\end{equation}         
was calculated for each frame $t$, where $n(i)$ is the number of clusters present in the $i$th frame, and $T$ is the total length of the simulation.
     
     In order to evaluate the thermodynamics of the finite-size effect MARTINI simulations, Tanford’s treatment of equilibrium micelle size distributions was applied. Tanford’s fundamental equation\cite{Tanford_1974} defines the mole fraction $X_s$ of aggregate size $s$
\begin{equation}
\textrm{ln}(X_s ) = –\frac{s \Delta G_s^0}{\textrm{RT}} + s\textrm{ln}(X_1) + \textrm{ln}(s)
\label{equation2}
\end{equation}
where $\Delta G_s^0$ is the free energy change of adding one surfactant to a micelle of size $s$ – 1, and $X_1$ is the mole fraction of monomers in the system. The value of $s$ at which the slope of ln($X_s$) is zero corresponds to the preferred aggregation number, $s$*, of the surfactant system. A micelle of size $s$* has the optimal balance of entropy and enthalpy, which mainly correspond to surfactant tail conformational distributions and head group solvation, respectively\cite{Tanford_1974}.

     Tanford also constructed a gaussian model that can be used to approximate the preferred aggregation number, $s$*, from a micelle size distribution which has the form
\begin{equation}
X_s = X_s^*\textrm{e}^{–a(s – s^* )^2}
\label{equation3}                                                   
\end{equation}
where $X_s$ is the mole fraction of micelles with $s$ members, $X_s$* is the mole fraction of micelles with $s$* members, and $a$ is a constant parameter\cite{Tanford_1974}.
 
 \subsection{G. Comparison with experimental SANS spectrum}
     The results of the all atom simulations were used to construct simulated SANS profiles through the use of the Debye scattering equation\cite{Pedersen_1997}. The Debye scattering equation solves for the scattering intensity defined in terms of the interatomic distances, $r_{ij}$, within each aggregate, such that 
\begin{equation}
\frac{ \textrm{I}(Q)}{\textrm{I}(Q_0)} = \frac{\sum_{i,j}b_ib_j^*\frac{\textrm{sin}(Qr_{ij})}{Qr_{ij}}}{\sum_{i,j}b_ib_j^*\frac{\textrm{sin}(Q_0r_{ij})}{Q_0r_{ij}}}
\label{equation4}
\end{equation}
where $b_i$ and $b_j$ are the scattering lengths\cite{Keller_1996} of atoms $i$ and $j$, and $Q$ is the scattering vector\cite{Dinnebier_Billinge_2019}. To make a direct comparison between the results of the all atom simulations and the experimental results of Pambou \latin{et al.}\cite{Pambou_Crewe_Yaseen_Padia_Rogers_Wang_Xu_Lu_2015}, the intensity vs $Q$ points for the fully deuterated sample of DPC at 100 mM and 295 K were extracted from Figure 4 using Engauge Digitizer Software\cite{Mitchell_Muftakhidinov_Winchen}. The scattering lengths used for each DPC atom in femtometers are 0.66 for ${}^{12}$C, 0.65 for ${}^{2}$H, 0.94 for ${}^{14}$N, 0.58 for ${}^{16}$O, and 0.50 for P\cite{Keller_1996}. 

     In addition, the radius of gyration ($R_g$) of the micelles was calculated from the distance distribution function\cite{Hammouda_2016,Hasko_Paradies_1980}, $\mathrm{p}(r)$, which is the probability function of interatomic distances, $r_{ij}$, within each micelle. $R_g^2$ is half of the second moment of the pair distance distribution function\cite{Hasko_Paradies_1980},
\begin{equation}
R^2_G = \frac{\int_{0}^{\infty}r^2\textrm{p}(r)\textrm{dr}}{2\int_{0}^{\infty}\textrm{p}(r)\textrm{dr}}
\label{equation5}
\end{equation}
or $\sqrt{3/5}$ times the mean radius of the pair distance distribution function\cite{Abel_Dupradeau_Marchi_2012} of an approximately spherical micelle. The radius of gyration was used as another way to compare the simulated and experimental DPC micelles.

\section{Results and discussion}

\subsection{A. Identifying the optimal CG model for DPC micelle self-assembly}

\begin{figure*}
\begin{center}
\includegraphics{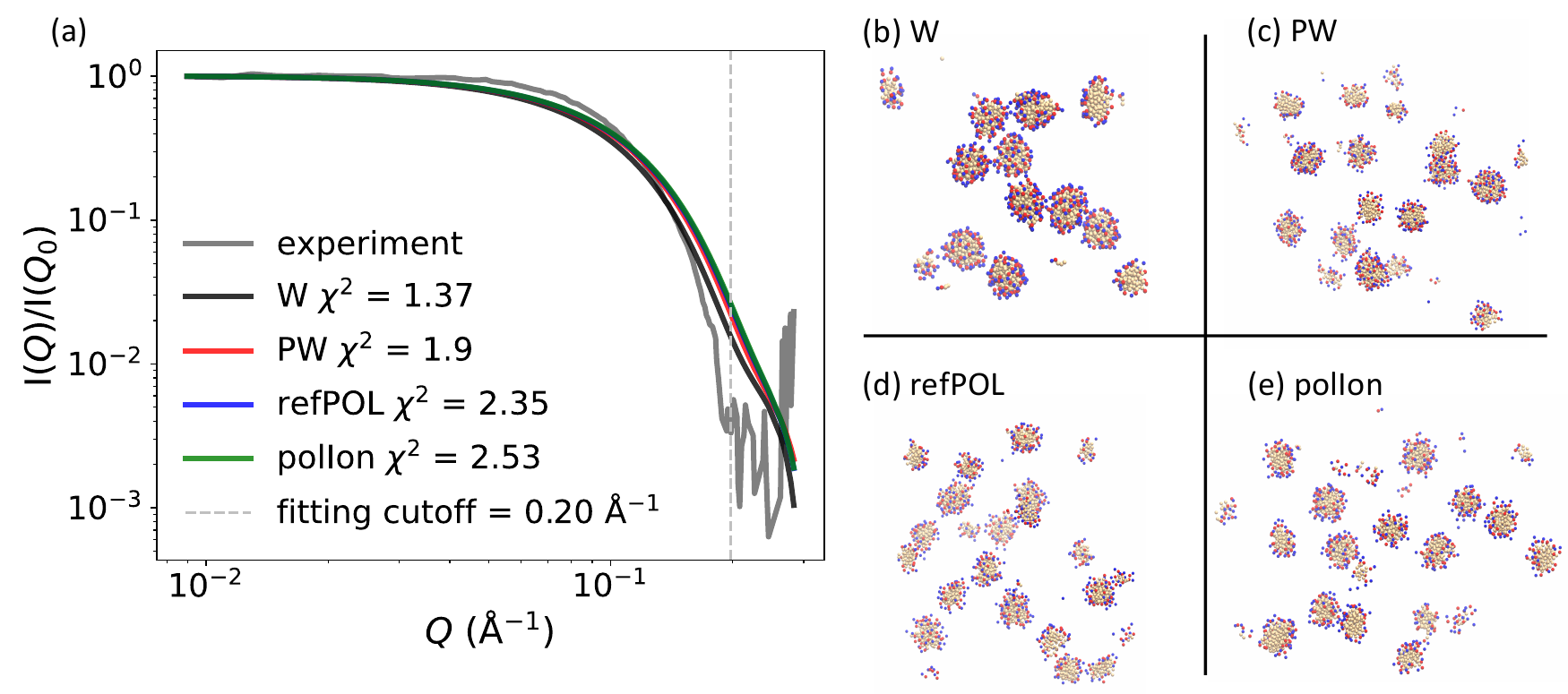}
\end{center}
\caption{(a) Normalized SANS spectra computed from DPC micelle data for each MARTINI simulation with solvent models W (black), PW (red), refPOL (blue), and polIon (green), respectively.  Experimental spectrum from Pambou et al. (grey)\cite{Pambou_Crewe_Yaseen_Padia_Rogers_Wang_Xu_Lu_2015} shows best agreement with DPC in the W water model. The dashed line is the cutoff $Q$ value for the $\chi^2$ analysis between each model and the experimental profile. Snapshots of the DPC micelle system with (b) W water model at 5.0 $\mu$s, (c) PW water model at 4.3 $\mu$s, (d) refPOL water model at 3.3 $\mu$s, and (e) polIon water model at 4.2 $\mu$s.}
\label{figure1}
\end{figure*} 

     As a way of determining which MARTINI solvent model is most appropriate for micelle self-assembly, the equilibrium distributions of micelles resulting from the large ($N$ = 500 DPC) simulations of each model were compared directly to the SANS profile produced by Pambou \latin{et al.}\cite{Pambou_Crewe_Yaseen_Padia_Rogers_Wang_Xu_Lu_2015} The  systems were considered to have reached equilibrium after 2.5 $\mu$s based on the convergence of the ergodic measure of the average of the number of clusters present (see SI). Because the MARTINI representation does not include every atom present in DPC, scattering lengths for each DPC bead were approximated based on the 157.63 fm deuterated head and 246.53 tail scattering lengths\cite{Pambou_Crewe_Yaseen_Padia_Rogers_Wang_Xu_Lu_2015} reported by Pambou \latin{et al.} so that the scattering lengths are 78.65 fm for each of the two head beads and 82.18 fm for each of the three tail beads. It was found from a $\chi^{2}$ analysis that the nonpolarizable W model yields the most realistic micelle distribution, with a $\chi^{2}$ of 1.37, while PW had 1.90, refPOL had 2.35, and polIon had 2.53 (Fig. \ref{figure1}), where 
\begin{equation}
\chi^2 = \frac{[\textrm{I}(Q)/\textrm{I}(Q_0)^{sim} \  – \ \textrm{I}(Q)/\textrm{I}(Q_0)^{exp}]^2}{\textrm{I}(Q)/\textrm{I}(Q_0)^{exp}} 
\label{equation6}
\end{equation}
for each $Q$ value from 0.0090 Å$^{-1}$ up to 0.20 Å$^{–1}$.

      The performances of the MARTINI solvent models were also evaluated based on which size distribution is most reasonable when compared to previous DPC micelle experiments. In order to quantify the micelle size distributions produced by each water model, Tanford's gaussian approximation (Eq. \ref{equation3}) was used to calculate preferred aggregation numbers ($s^*$). The gaussian function was applied to the size distributions which contained data from 6 replicate simulations of each water model (see SI). It was found that the W model yielded $s^*$ =  44.2 $\pm$ 7.0, PW = 36.2 $\pm$ 9.9, refPOL = 31.8 $\pm$ 7.7, and polIon = 33.0 $\pm$ 7.5 surfactants. It is fortuitous that the W model is also the most computationally affordable of the MARTINI solvents, with production speeds 4 times that of the other models.

\subsection{B. Finite-size effects on DPC self-assembly using CG models}
\begin{figure*}[!htb]
\begin{center}
\includegraphics{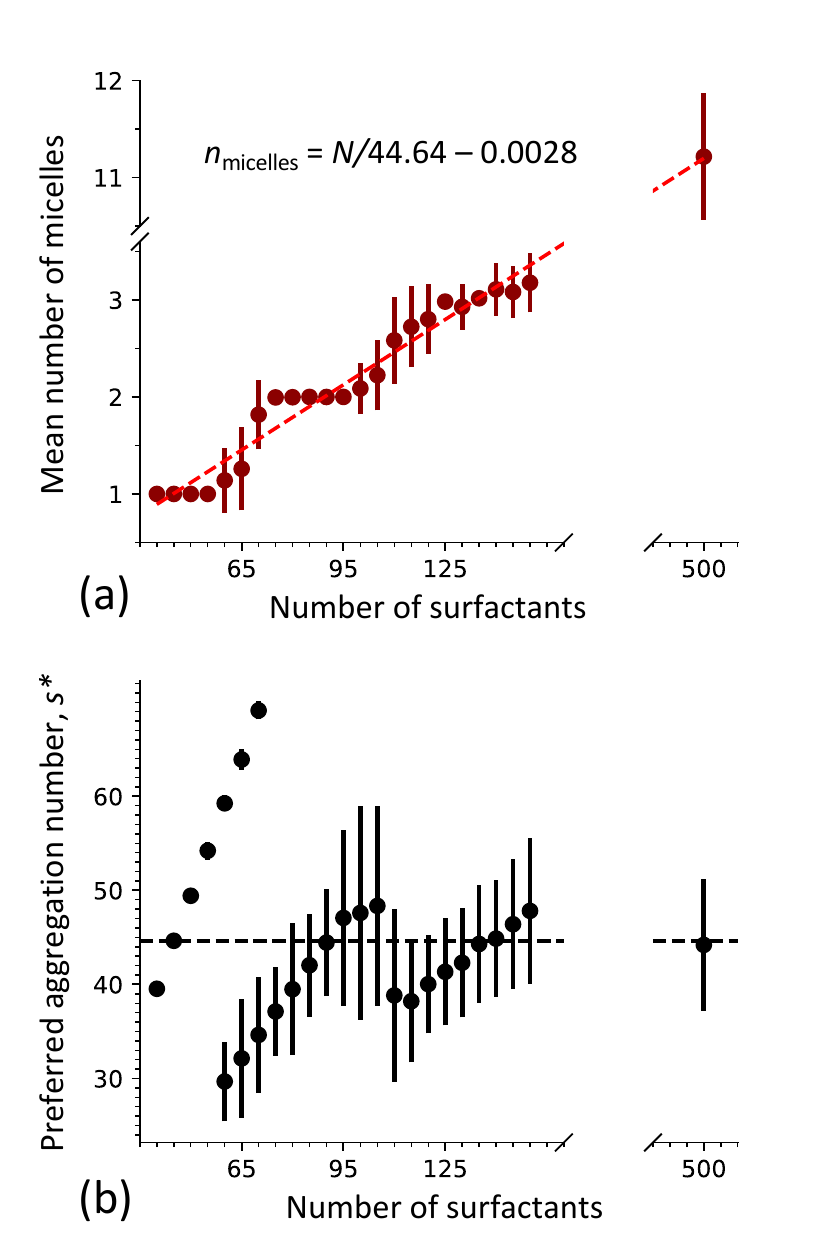}
\end{center}
\caption{(a) The mean number of micelles in each system as a function of the total number of surfactant molecules in the system ($N$). Error bars represent the standard deviation across replicates. Linear fit to data defined as: $n_{\mathrm{micelles}}$ = $N$/44.64 – 0.0028, which corresponds to $s^{\mathrm{thermo}}$ = 44.64 and $c$ = – 0.0028 (Eq. \ref{equation7}). (b) Preferred aggregation number, $s^*$, for each system as a function of total number of surfactants ($N$). The $s^*$ values were obtained by Tanford's gaussian approximation for micelle size distributions (Eq. \ref{equation3}). The horizontal line corresponds to $s^{\mathrm{thermo}}$ = 44.64. System sizes $N$ = 60, 65, and 70 have two $s^*$ values due to their bimodal size distributions. Error bars represent the standard deviation of the fitted normal distributions to the micelle size distributions.}  
\label{figure2}
\end{figure*}

    The initial characterization of finite-size artefacts was completed by computing the average number of micelles, the preferred aggregation number ($s^*$) for each system size, and the aggregation number expected in the thermodynamic limit, $s^{\mathrm{thermo}}$ (Fig. \ref{figure2}). The micelle equilibrium was considered to be achieved after 1.5 $\mu$s for systems N = 40 to 100  and 3.0 $\mu$s for systems N = 105 to 150. Equilibrium was determined by the time series analysis of the convergence of the ergodic measure based on the time averaged number of clusters per system (see SI). 
    
    For the purposes of data analysis, a micelle is defined as a cluster with more than 10 DPC. Qualitatively, the plot of the number of micelles ($n_{\mathrm{micelles}}$) as a function of the total number of surfactants ($N$) in the systems reveals the finite-size behavior previously proposed by Kindt\cite{Kindt_2013}. We see characteristic plateaus associated with integer numbers of micelles for smaller $N$ systems and convergence to an approximately linear relationship at larger $N$ (Fig. \ref{figure2}). Kindt demonstrated that in the thermodynamic limit of large $N$, the relationship between the $N$ and $n_{\mathrm{micelles}}$ is linear\cite{Kindt_2013}. As a way of approximating this correlation, a linear fit was obtained,
\begin{equation}
n_{\textrm{micelles}} = \frac{N}{s^{\textrm{thermo}}} + c
\label{equation7}
\end{equation}
where the slope, $1/s^{\textrm{thermo}}$ corresponds to the number of micelles per surfactant in the thermodynamic limit, and the inverse of the slope represents the number of surfactants per micelle at this limit. The number of surfactants per micelle in the thermodynamic limit is a special aggregation number, $s^{\mathrm{thermo}}$, which was found to be 44.64, with a small correction constant, $c$ = –0.0028. As expected, systems $N$ = 45, 90, and 135, which contain multiples of $s^{\mathrm{thermo}}$, fall on the line defined by Eq. \ref{equation7} (Fig. \ref{figure2}). We therefore suggest that these system sizes are less impacted by finite-size effects because they contain appropriate numbers of surfactants to form 1, 2, and 3 micelles of size $s^{\mathrm{thermo}}$. The relationship between each $s^*$ micelle size and the total number of surfactants is also in agreement with previous predictions by Kindt\cite{Kindt_2013}. The plot of the $s^*$ values presents a damped oscillation and converges to the $s^{\mathrm{thermo}}$ $\approx$ 45 number of surfactants (Fig. \ref{figure2}).   

\begin{figure*}[!htb]
 \begin{center}
\includegraphics{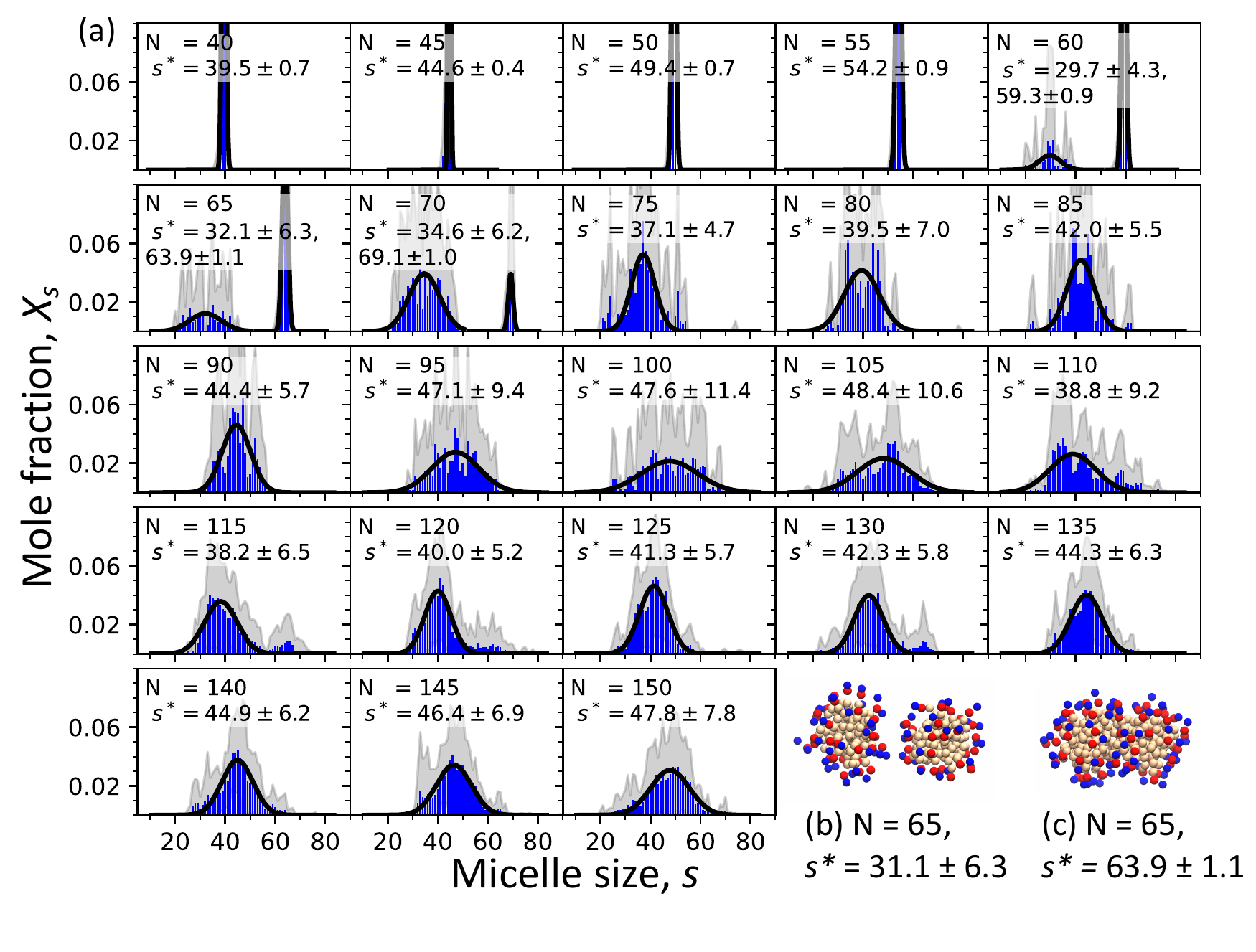}
\end{center}
\caption{(a) Micelle size distributions for each system (blue bars) measured as the mole fraction $X_s$ of aggregates of size $s$ with standard deviation across replicates (grey fill). Fits to a gaussian distribution centered at $s^*$ defined by Tanford (black line). Derived most probable $s^*$ values reported with standard deviation of the normal distribution. Snapshots depicting equilibrium micelle ensembles from systems (b) $N$ = 65 at 1.82 $\mu$s and (c) $N$ = 70 at 1.7 $\mu$s.}
\label{figure3}
\end{figure*}

     In order to derive the preferred aggregation number, $s^*$, for each system size from the equilibrium micelle distribution, we consider Tanford’s gaussian approximation for modeling micelle size distributions (Eq. \ref{equation3})\cite{Tanford_1974}. After fitting the gaussian function to selected regions of the computed mole fraction distribution of each system size, the finite-size effect is again apparent (Fig. \ref{figure3}). Systems of 60, 65, and 70 DPC exhibit unique, bimodal size distributions due to their relatively small finite-size, and consequently have two $s^*$ values. The $s^*$ values for each system are summarized in Table \ref{table1}. A notable result in the large $N$ region is that past $N$ = 135, the $s^*$ value increases. When $N$ = 135, $s^*$ = 44.3. If we assume that the previously calculated thermodynamic limit $s^*$ value of 44.6 is accurate, we find that increasing the total number of surfactants from $N$ = 135 to $N$ = 150 brings us further away from our expected result. When $N$ = 150, the gaussian fit leads to an $s^*$ value of 47.8. Therefore, we see that even at the system size of $N$ = 150, there remains, although diminished, an artefact of the finite-size effect. After applying the same gaussian fitting to the $N$ = 500 system described in the previous section, which had the same concentration and solvent model as the finite-size effect simulations, we obtain $s^*$ = 44.2 $\pm$ 7.0 (see SI). In this large $N$ system, we find that the preferred aggregation number is similar to that observed in the smaller $N$ = 135 system, which further validates our method of calculating $s^*$ values using relatively small system sizes.
 
\begin{table*}
\begin{center}
\includegraphics{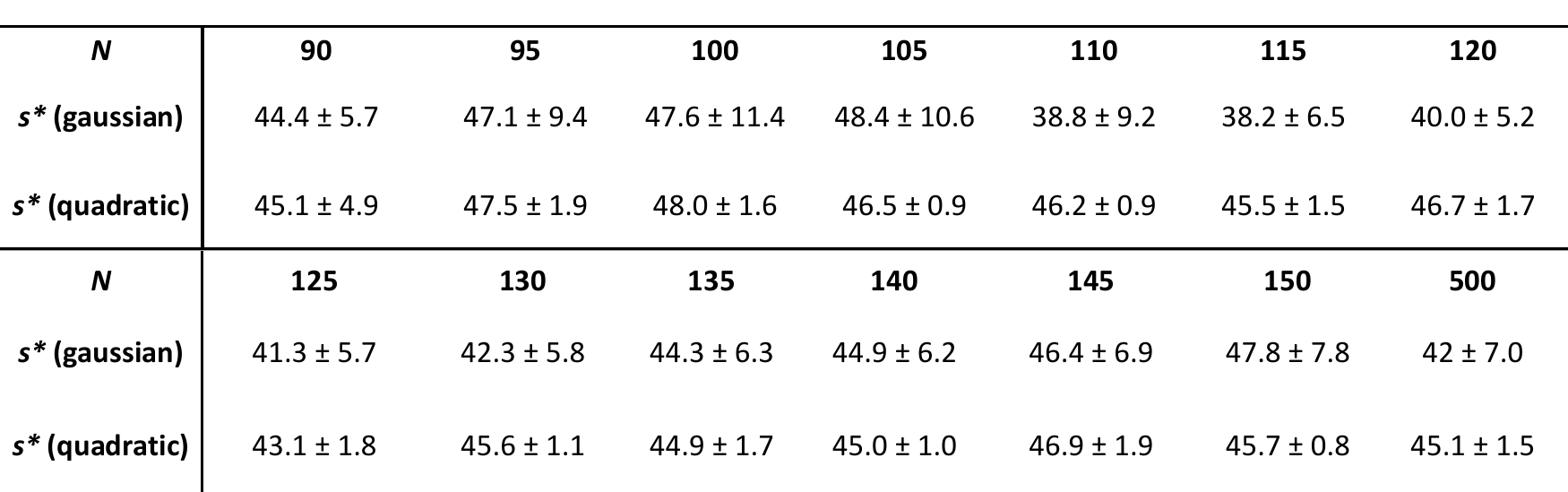}
\end{center}
\caption{Summary of $s^*$ values calculated by gaussian fitting to $X_s$ and quadratic fitting to ln($X_s$)}
\label{table1}
\end{table*}
     An alternative way of computing $s^*$ from the finite-size simulations makes use of Tanford’s fundamental micelle equation (Eq. \ref{equation2})\cite{Tanford_1974}. The value of $s$ which maximizes the equation corresponds to the $s^*$ value of the size distribution. As $X_s$ can be modeled as a gaussian function, ln($X_s$) can be modeled with a quadratic function that can be used to identify $s^*$ for each system size. The quadratic function was successful for fitting to the medium and large $N$ systems, but did not work well for small $N$ systems due to both non-gaussian and bimodal size distributions (see SI). The obtained $s^*$ values for each system are summarized in Table \ref{table1}. 
     
\begin{figure*}[!htb]
\begin{center}
\includegraphics{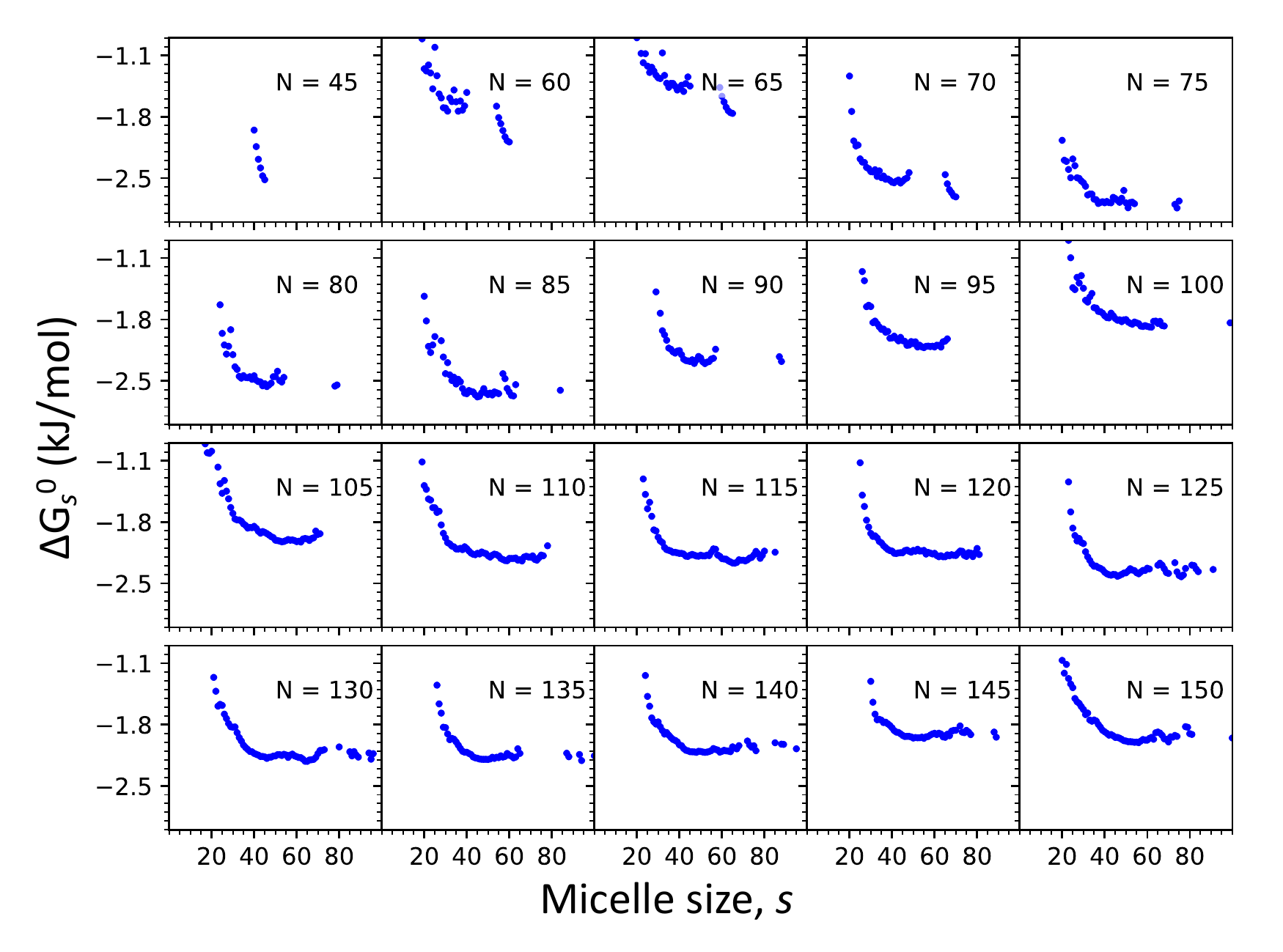}
\end{center}
\caption{Calculated $\Delta G_s^0$ values for each system over the micelle size distribution, where $\Delta G_s^0$ is the free energy change for adding a surfactant to a pre-existing micelle of size $s$ in the Tanford model. The plateau observed at large $s$ for $N$ $>$ 135 suggests $\Delta G_s^0$ $\approx$ –2.0 kJ/mol.}
\label{figure4}
\end{figure*}

     To further evaluate the thermodynamics of micellization as observed in the finite-size simulations, we can invoke Tanford’s micelle equation (Eq. \ref{equation2}) to derive $\Delta G_s^0$ from the distribution of the mole fraction, $X_s$\cite{Tanford_1974}. If we consider each micelle of size $s$ as being in equilibrium with monomers in each system, we expect that 
\begin{equation}
\Delta G_s^0 = –\frac{\textrm{RT}}{s}\textrm{ln}(K_s)    
\label{equation8}                                    
\end{equation}
where $K_s$ = $X_s$ / $sX_1$  is the equilibrium constant for the formation of a micelle with $s$ members from $s$ surfactants. The value of $\Delta G_s^0$ corresponds to the free energy change\cite{Tanford_1974} resulting from adding one more surfactant to a micelle of size $s$.  The resulting $\Delta G_s^0$ values are on the order of –1 kJ/mol (Fig. \ref{figure4}). In the bimodal region of the free energy plots, $N$ = 60, 65, and 70, we see an interesting trend in the $\Delta G_s^0$ values, which have double wells around two separate preferred aggregation numbers. The finite-size effect on the free energy of micellization is especially apparent for these systems. 

\subsection{C. Back-mapping MARTINI representations to CHARMM36 all-atom representations leads to decrease in micelle volume }  
     
\begin{figure*}[!htb]
\begin{center}
\includegraphics{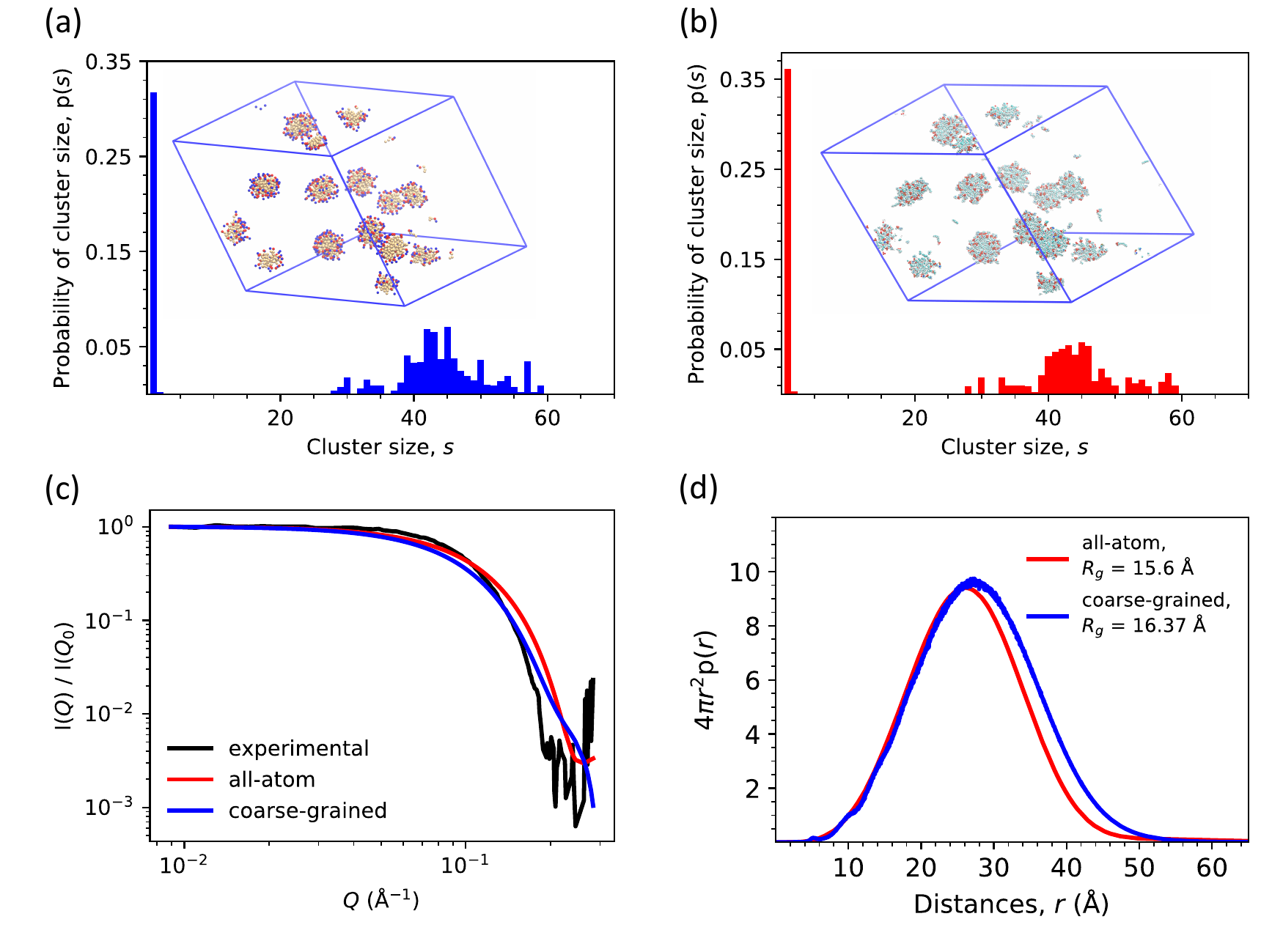}
\end{center}
\caption{(a) Cluster size distribution of MARTINI W system, averaged over the last 20 ns of the 6 replicates. Superimposed is the first frame before back-mapping. (b) Cluster size distribution of the back-mapped AA system, averaged over the first 20 ns, omitting the first nanosecond. Superimposed is the first frame after back-mapping. (c) SANS profiles constructed using the Debye scattering equation (Eq. \ref{equation4}), for the AA (red), MARTINI (blue), and experimental\cite{Pambou_Crewe_Yaseen_Padia_Rogers_Wang_Xu_Lu_2015} (black) systems. (d) Distance distribution function, 4$\pi$$r^2$p($r$), and corresponding radius of gyration calculated using the second moment of the distribution (Eq. \ref{equation5}) for the equilibrium ensemble of the MARTINI (blue) and first 20 ns, omitting the first ns, of the AA (red) systems.}
\label{figure5}
\end{figure*}

     The finite-size effects on self-assembly could possibly be avoided by implementing a multiscale approach; first simulating a large, computationally affordable MARTINI simulation, and then back-mapping the result to an all atom representation for better comparison with experimental data.  In the interest of testing the applicability of this method, the DPC positions at 5200 ns from the large $N$ = 500 DPC MARTINI W systems in section A. were back-mapped to CHARMM36 AA representations using the backward.py method\cite{Wassenaar_2014}. The new AA systems were simulated for 20 ns until the time series of the mean radius of gyration ($R_g$) was observed to converge (see SI). The systems were analyzed to produce a micelle size distribution, SANS profile, and pair distance distribution function (Fig. \ref{figure5}). The first nanosecond was omitted from these calculations because there was a substantial collapse in $R_g$ after back-mapping from MARTINI to CHARMM36 (see SI). 
     
     The radius of gyration of the AA micelles was found to be 15.6 Å, in contrast with the MARTINI radius of gyration of 16.37 Å. The decrease in $R_g$ after the back-mapping was performed is evidence, however, that the MARTINI surfactants occupy a larger volume within a smaller aggregation number, compared to the AA model. Our results with the MARTINI model are similar to the findings of Abel \latin{et al.}, who reported a $R_g$ of 16.8 Å for an AA DPC micelle of size $s$ = 54 with CHARMM36\cite{Abel_Dupradeau_Marchi_2012}. In addition, the all-atom simulations of Faramarzi \latin{et al.} yielded a similar $R_g$\cite{Faramarzi_Mertz_Bonnett_Scaggs_Hoffmaster_Grodi_Harvey_2017}. The preferred aggregation number of the MARTINI micelles, $s^*$ = 44.2 $\pm$ 7.0 (see SI), is considerably smaller than Abel's micelle. This $s^*$ value, however, is in agreement with the MARTINI self-assembly simulations of Sanders \latin{et al.}, which resulted in a micelle size distribution centered around 45 surfactants\cite{Sanders_Panagiotopoulos_2010}. This insight explains why the radius of gyration decreases after back-mapping, and also provides evidence that the MARTINI micelles reliably predict the preferred volume, but underestimate the preferred aggregation number because the DPC molecule itself is too large in this model. This finding is remarkable considering the major computational improvements afforded by the MARTINI model, which sacrifices accuracy in aggregation number, but does not compromise the expected $R_g$ of the micelles.
     
     At the concentration of 100 mM (used throughout our simulations), Pambou \latin{et al.} found a radius of gyration of 21.45 Å, based on the reported core radius of 19.6 Å and shell radius of 7.8 Å\cite{Pambou_Crewe_Yaseen_Padia_Rogers_Wang_Xu_Lu_2015}, and assuming spherical micelles for the purpose of approximation. In addition, the SAXS-derived $R_g$ for DPC micelles was found by Oliver et al. to be 34.5 $\pm$ 0.08 Å at 77 mM surfactant concentration\cite{Oliver_Lipfert_Fox_Lo_Doniach_Columbus_2013}. As noted by Faramarzi \latin{et al.}, in the case of SAXS-derived $R_g$, the hydration shell of the experimental micelle results in a larger value than that predicted by MD\cite{Faramarzi_Mertz_Bonnett_Scaggs_Hoffmaster_Grodi_Harvey_2017}. In the ultracentrifugation study, Lauterwein \latin{et al.} found a diameter of 47 Å at 20 mM, which converts to $R_g$ = 18.2 Å under the spherical assumption\cite{Lauterwein_Bosch_Brown_Wuthrich_1979}. Similarly, the NMR studies of Kallick \latin{et al.} reported a hydrodynamic radius of 18.65 $\pm$ 0.3 Å at 228 mM\cite{Kallick_Deborah_A_Tessmer_Michael_R_Watts_Charles_R_Li_1995}. We see much better agreement between the simulation-derived $R_g$ values and the ultracentrifugation and NMR experimentally derived values. These variation in $R_g$ are summarized in Table \ref{table2}.
\begin{table*}
\begin{center}
\includegraphics{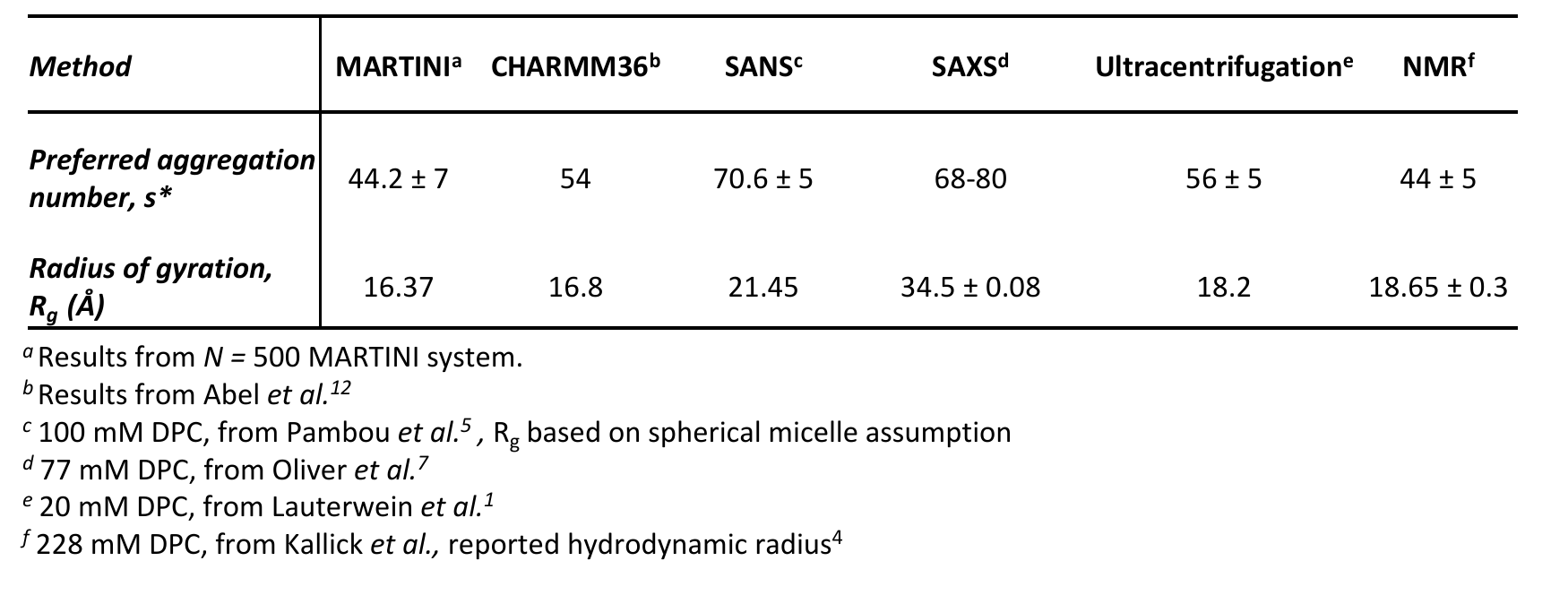}
\end{center}
\caption{Summary of $s^*$ values calculated by gaussian fitting to $X_s$ and quadratic fitting to ln($X_s$)}
\label{table2}
\end{table*}      

    Considering the experimental and simulated SANS profiles (Fig. \ref{figure5}), it is qualitatively evident that the experimental (black) and CG micelles (blue) have larger volumes, and the AA micelles (red) have smaller volumes. Furthermore, Pambou \latin{et al.} reported an aggregation number of 70.6 $\pm$ 5 surfactants\cite{Pambou_Crewe_Yaseen_Padia_Rogers_Wang_Xu_Lu_2015}. This aggregation number is not only larger than the value derived from the MARTINI simulations, but also larger than Lauterwein \latin{et al.}'s ultracentrifugation and DLS results\cite{Lauterwein_Bosch_Brown_Wuthrich_1979} of 56 $\pm$ 5 and Kallick \latin{et al.}'s NMR results\cite{Kallick_Deborah_A_Tessmer_Michael_R_Watts_Charles_R_Li_1995} of 44 $\pm$ 5 surfactants. These variations in aggregation number are summarized in Table \ref{table2}. The bulkiness of the MARTINI representations does not allow for an accurate number of surfactants to aggregate into a preferred micelle size. Due to this limitation of the MARTINI model, the large $N$, long timescale MARTINI simulations and subsequent back-mapping to AA, while effective in avoiding finite-size effects, is not adequate for reproducing experimental micelle results.  
    
\section{Conclusion}
     We performed molecular dynamics simulations of DPC micelle self-assembly to equilibrium at 100 mM surfactant concentration using four different MARTINI 2 solvent models and evaluated the equilibrium size distributions by direct comparison with an experimental SANS spectrum. We determined that the widely-used MARTINI 2 non-polarizable water produces SANS spectra most similar to experiment. We studied finite-size effects on DPC micelle simulations in 40- to 150-DPC systems. We observed damped oscillations in both the number of micelles and in the preferred aggregation number ($s^*$) of each system as a function of the number of surfactants in the system about the value of $s^*$ in the thermodynamic limit. We observe that these damped oscillations mostly converge to $s^{\mathrm{thermo}}$ once the system size exceeds 3 times the value of $s^{\mathrm{thermo}}$. This observation suggests that reasonably accurate micelle simulations may be performed by using three or more times a guessed $s^{\mathrm{thermo}}$ number of surfactants in a simulation, assuming that this value can be estimated. In addition to damped oscillations in micelle number and size as a function of system size, we also observed bimodal distributions of micelle sizes in systems of 60-, 65- and 70-DPC.     
     
     As a way of testing the accuracy of a multiscale approach to micelle self-assembly, large 500-DPC MARTINI micelle configurations at equilibrium were back-mapped to AA CHARMM36 representations. The aggregation numbers, radii of gyration, and SANS profiles were compared between models and with experiment. It was found that the MARTINI model is reliable for achieving the proper radius of gyration, but, due to the large volume of each MARTINI DPC molecule, underestimates the preferred aggregation number.
      
     These results provide a clear prescription for the accurate modeling of DPC micelle self-assembly in terms of the choice of model, minimum system size, and methods of analysis for comparison with experiment. The trade-offs between models and the feasibility of a multiscale approach to the problem of determining equilibrium micelle size distributions are elucidated. It is clear that the MARTINI model is accurate for determining the volume of DPC micelles, and was reliable for fully assessing the finite-size effect and deriving the thermodynamic aggregation number for a generic surfactant. For determining a more accurate aggregation number, however, an AA model must be employed, at the expense of longer computational time. We posit that the methods described here for determining the minimum system size for MARTINI DPC can be repeated and generalized for a wide variety of surfactants which may be of research interest. 

\bibliography{version5}

\newpage
\begin{center}
\textbf{\large Supporting Information: Finite-size effects and optimal system sizes in simulations of surfactant micelle self-assembly}
\end{center}
\setcounter{figure}{0}
\makeatletter
\renewcommand{\thefigure}{S\arabic{figure}} 

\begin{figure}
\includegraphics{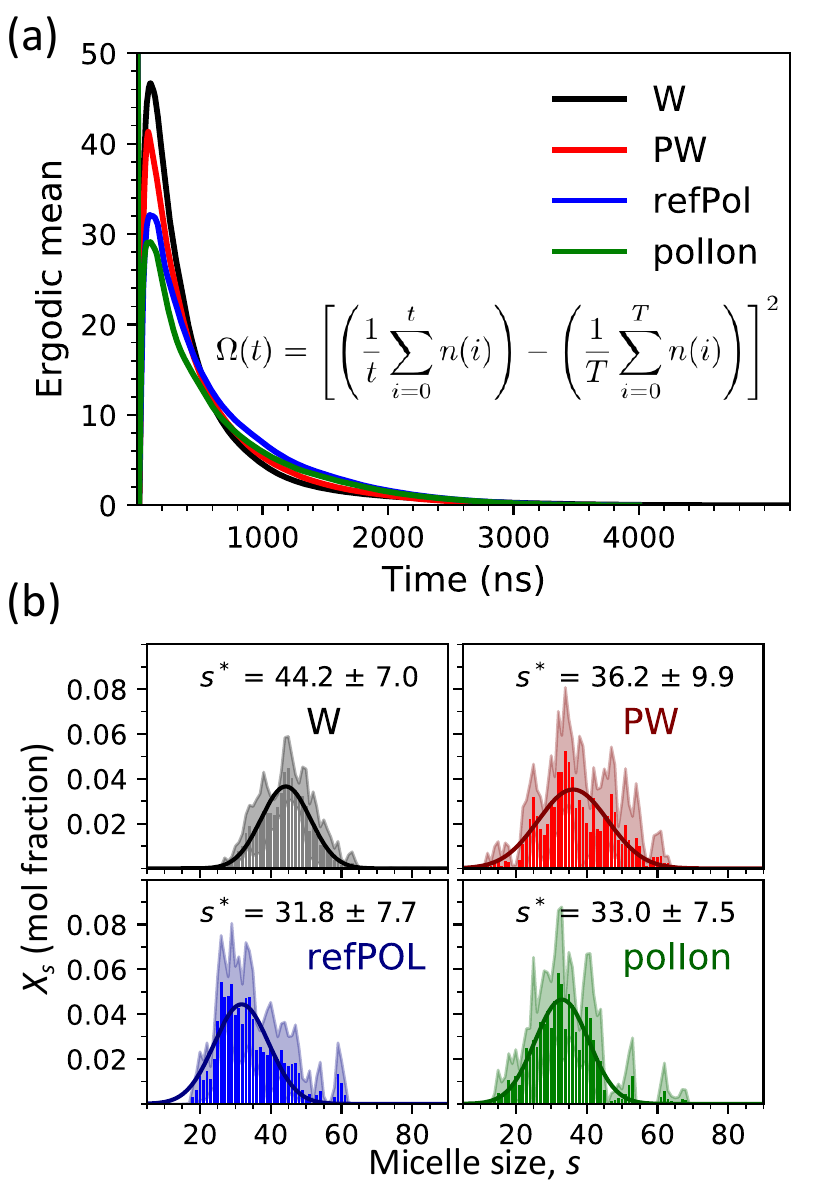}
\caption{(a) Ergodic time series measure of the average number of clusters up to time $t$, minus the average number of clusters over the entire trajectory. The $N$ = 500 DPC MARTINI systems are W (black), PW (red), refPol (blue), polIon(green). (b) Mole fraction size distributions for each of the $N$ = 500 MARTINI systems. The Tanford gaussian approximation is applied so that the preferred aggregation number, $s^*$, is derived.}  
\end{figure}  
  
\begin{figure}
\includegraphics{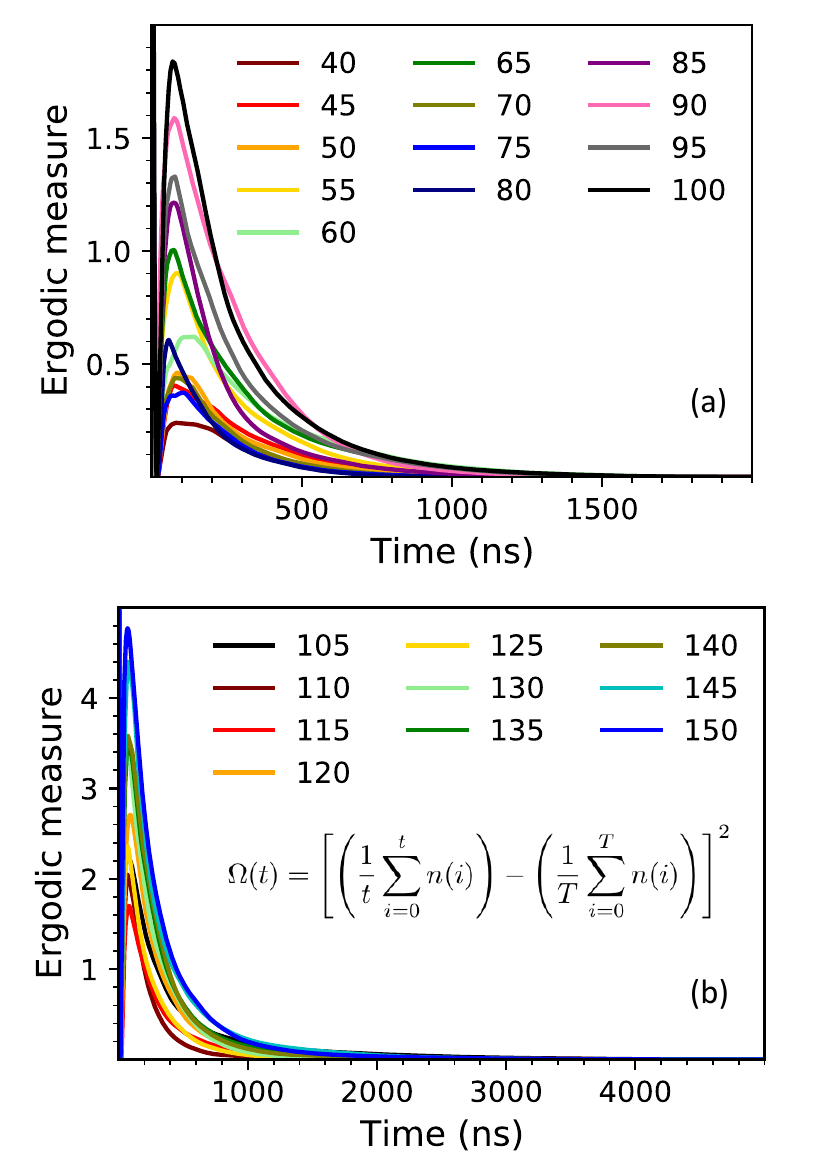}
\caption{Ergodic time series measure of the average number of clusters up to time $t$, minus the average number of clusters over the entire trajectory for the finite-size effect MARTINI systems. Each color corresponds to a different system size.}  
\end{figure}  

\begin{figure}
\includegraphics{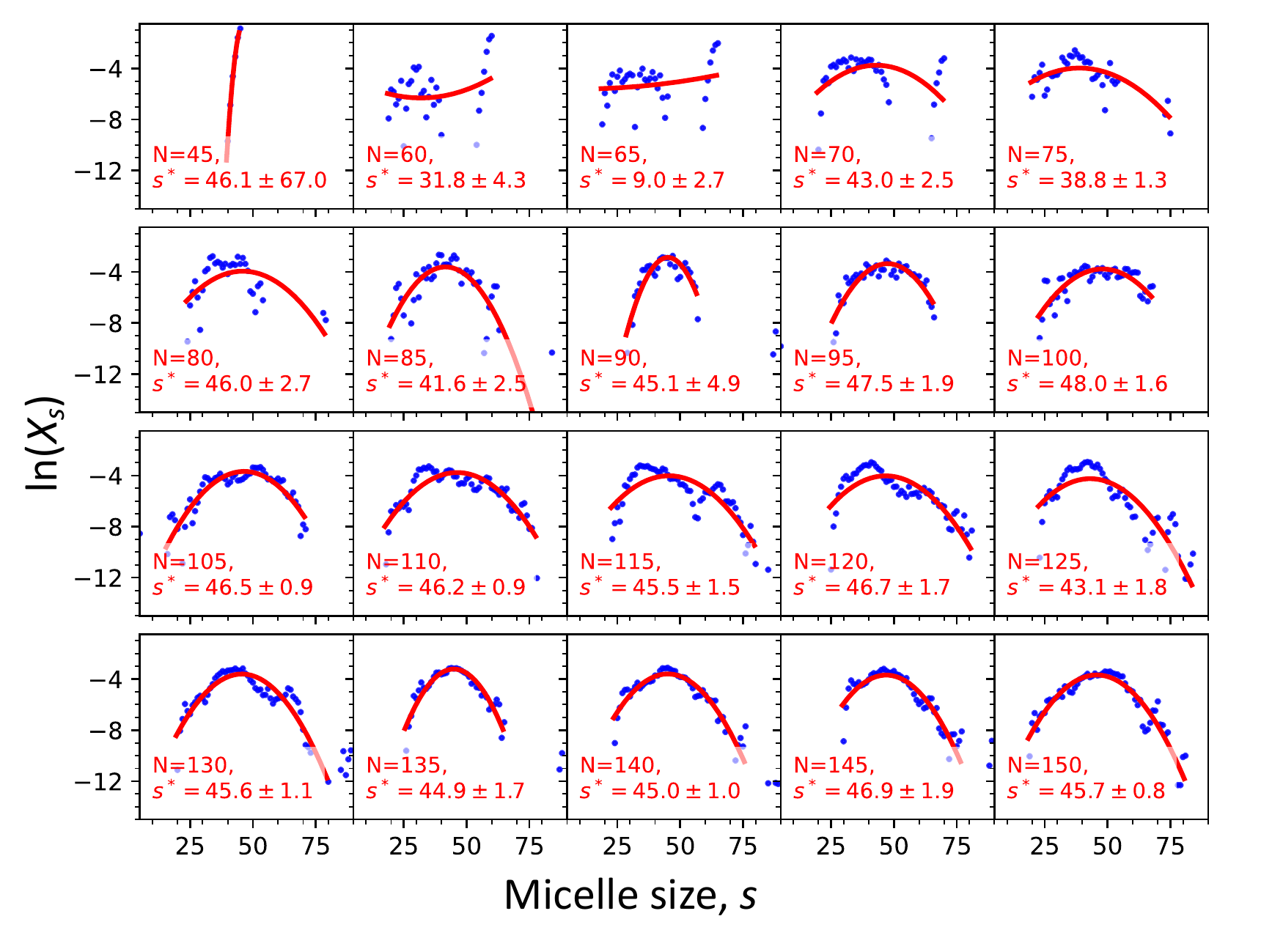}
\caption{Natural logarithm of the mole fraction of each cluster size for each finite-size effect system micelle size distribution. Tanford's fundamental micelle equation is applied to derive the $s^*$ value for each system size. The $s^*$ value is the maximum of the quadratic fit to the points.}  
\end{figure}  

\begin{figure}
\includegraphics{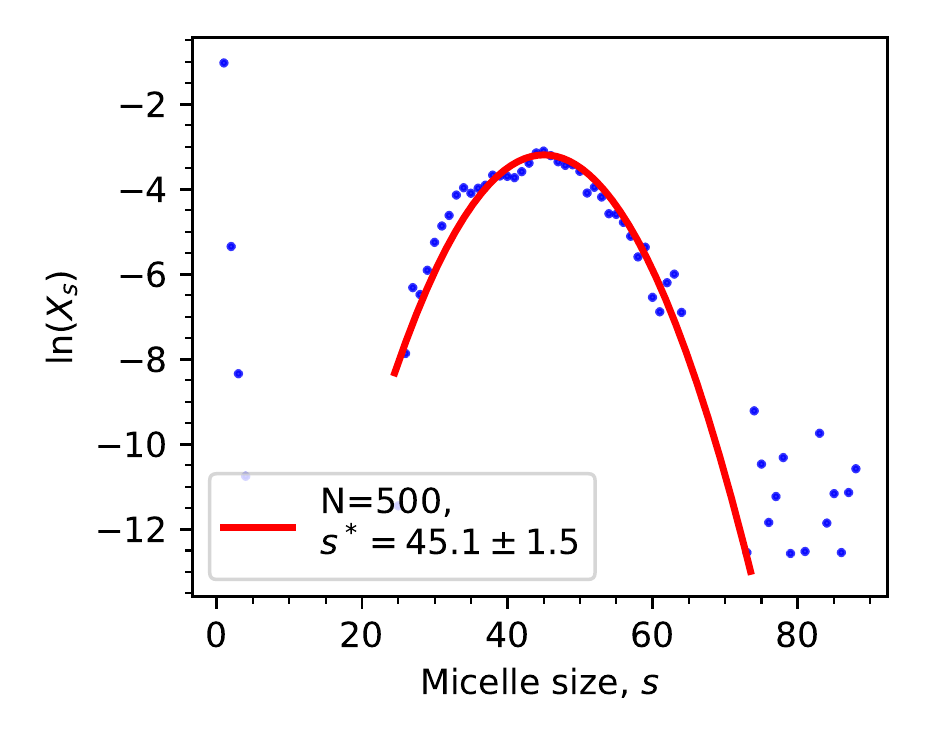}
\caption{Natural logarithm of the mole fraction of each cluster size for the $N$ = 500 system. Tanford's fundamental micelle equation is applied to derive the $s^*$ value for each system size. The $s^*$ value is the maximum of the quadratic fit to the points.}  
\end{figure}  

\begin{figure}
\includegraphics{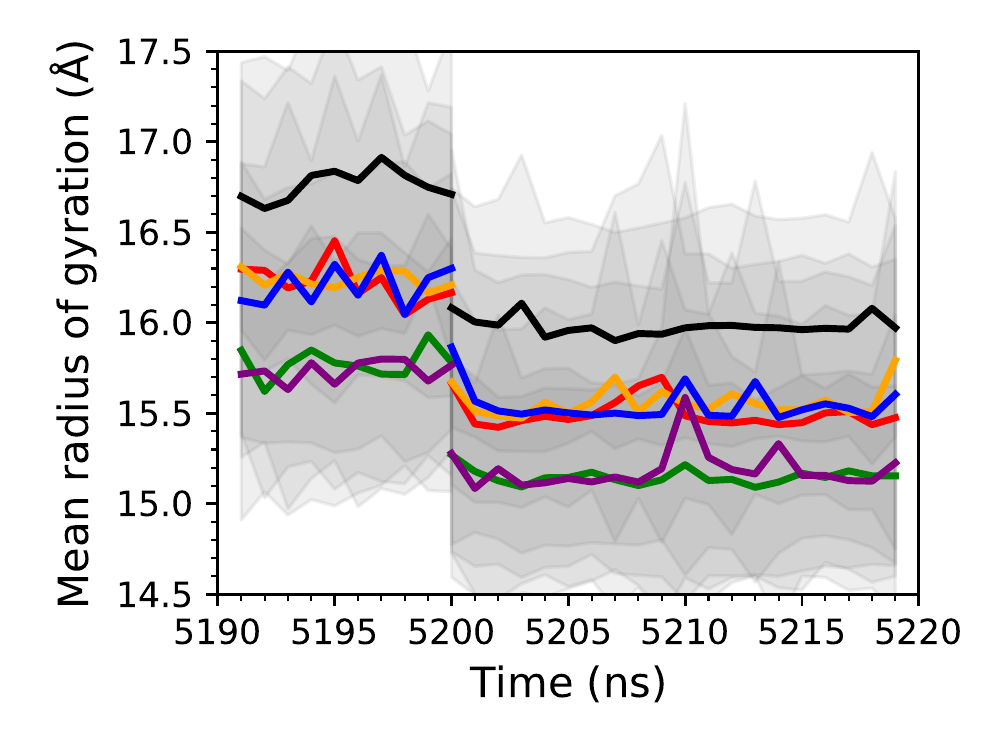}
\caption{Time series of the mean radius of gyration of micelles with more than 10 surfactants for each of the six replicates of the back-mapped systems. The back-mapping to AA representations was applied at 5200 ns. After 5200 ns, the radius of gyration is averaged over each ns from ten 0.1 ns sized data points.}  
\end{figure}  

\end{spacing}
\end{onecolumn}


\end{document}